\documentclass[a4paper]{IEEEtran}
\IEEEoverridecommandlockouts
\usepackage{cite}
\usepackage{amsmath,amssymb,amsfonts}

\usepackage{graphicx}
\usepackage{color}
\usepackage{subfig}
\usepackage{stackengine}
\usepackage{comment}
\usepackage{ulem}
\usepackage{tabularx}
\usepackage{bm}
\usepackage[font=scriptsize]{caption}
\usepackage[left=1.53cm,right=1.53cm, top=1.778cm, bottom=4.6cm]{geometry}

\usepackage{algorithm} 
\usepackage{algorithmic}  

\usepackage[linesnumbered, ruled, algo2e]{algorithm2e}
\usepackage[colorlinks, citecolor=blue]{hyperref} 

\usepackage[flushleft]{threeparttable} 
\usepackage{booktabs}

\usepackage{textcomp}
\usepackage{xcolor}

\def\BibTeX{{\rm B\kern-.05em{\sc i\kern-.025em b}\kern-.08em
    T\kern-.1667em\lower.7ex\hbox{E}\kern-.125emX}}
    
\usepackage{xspace}
\usepackage{ifthen}
\usepackage[inline]{enumitem}

\newboolean{showcomments}
\setboolean{showcomments}{true}
\ifthenelse{\boolean{showcomments}}
{
}

\usepackage{subfig}
\usepackage{fancyhdr}

\newcommand*{\affmark}[1][*]{\textsuperscript{#1}}

\begin{document}

\title{Hierarchical Deep Q-Learning Based Handover in Wireless Networks with Dual Connectivity}
\author{\small\IEEEauthorblockN{ Pedro Enrique Iturria-Rivera\affmark[1], \IEEEmembership{\small Student Member,~IEEE},  Medhat Elsayed\affmark[2], Majid Bavand\affmark[2], Raimundas Gaigalas\affmark[2],\\ Steve Furr\affmark[2] and Melike Erol-Kantarci\affmark[1], \IEEEmembership{\small Senior Member,~IEEE}}\\
\IEEEauthorblockA{\affmark[1]\textit{School of Electrical Engineering and Computer Science, University of Ottawa, Ottawa, Canada}}\\  \affmark[2]\textit{Ericsson Inc., Ottawa, Canada}\\
Emails:\{pitur008, melike.erolkantarci\}@uottawa.ca, \{medhat.elsayed, majid.bavand, raimundas.gaigalas, steve.furr\}@ericsson.com\vspace{-3em} }

\maketitle

\begin{abstract}
 5G New Radio proposes the usage of frequencies above 10 GHz to speed up LTE's existent maximum data rates. However, the effective size of 5G antennas and consequently its repercussions in the signal degradation in urban scenarios makes it a challenge to maintain stable coverage and connectivity. In order to obtain the best from both technologies, recent dual connectivity solutions have proved their capabilities to improve performance when compared with coexistent standalone 5G and 4G technologies. Reinforcement learning (RL) has shown its huge potential in wireless scenarios where parameter learning is required given the dynamic nature of such context. In this paper, we propose two reinforcement learning algorithms: a single agent RL algorithm named Clipped Double Q-Learning (CDQL) and a hierarchical Deep Q-Learning (HiDQL) to improve Multiple Radio Access Technology (multi-RAT) dual-connectivity handover. We compare our proposal with two baselines: a fixed parameter and a dynamic parameter solution. Simulation results reveal significant improvements in terms of latency with a gain of 47.6\% and 26.1\% for Digital-Analog beamforming (BF),  17.1\% and 21.6\% for Hybrid-Analog BF, and 24.7\% and 39\% for Analog-Analog BF when comparing the RL-schemes HiDQL and CDQL with the with the existent solutions,  HiDQL presented a slower convergence time, however obtained a more optimal solution than CDQL. Additionally, we foresee the advantages of utilizing context-information as geo-location of the UEs to reduce the beam exploration sector, and thus improving further multi-RAT handover latency results.

\end{abstract}

\vspace{-0.3cm}

\section{Introduction}
With the deployment of 5${th}$ Generation New Radio (5G)  and the existing 4${th}$ Generation LTE (4G) technologies, mobile users with LTE or 5G capabilities must be able to seamlessly adapt to the dual existent infrastructure. Back in 2013, the 3rd Generation Partnership Project (3GPP) proposed dual connectivity (DC) architectures in \cite{3GPP} that allowed master eNodeBs (eNBs) and secondary eNodeBs to share partially  their IP layer in a dual connection manner with the purpose of maximizing network performance. A more recent technical report \cite{3GPPTS37.3402019} proposed a generalization of multi-radio dual connectivity architectures to support LTE and 5G users. Dual connection architectures take advantage of the technologies involved to improve key performance indicators (KPIs) of interest. Concurrently, the stringent requirements of diverse services in terms of latency and throughput, such as Ultra-Reliable and Low Latency Communications (URLLC) and enhanced Mobile Broadband (eMMB) have demanded a more optimized parameter tuning in any wireless mechanism utilized.



Reinforcement Learning (RL) techniques have been widely recognized for its effectiveness in the autonomous learning context. More specifically, RL has sparked the wireless network community's attention\cite{Elsayed2019} since optimized parameter learning has become a challenge in such dynamic environments. 
\begin{figure}
\center
  \includegraphics[scale=0.16]{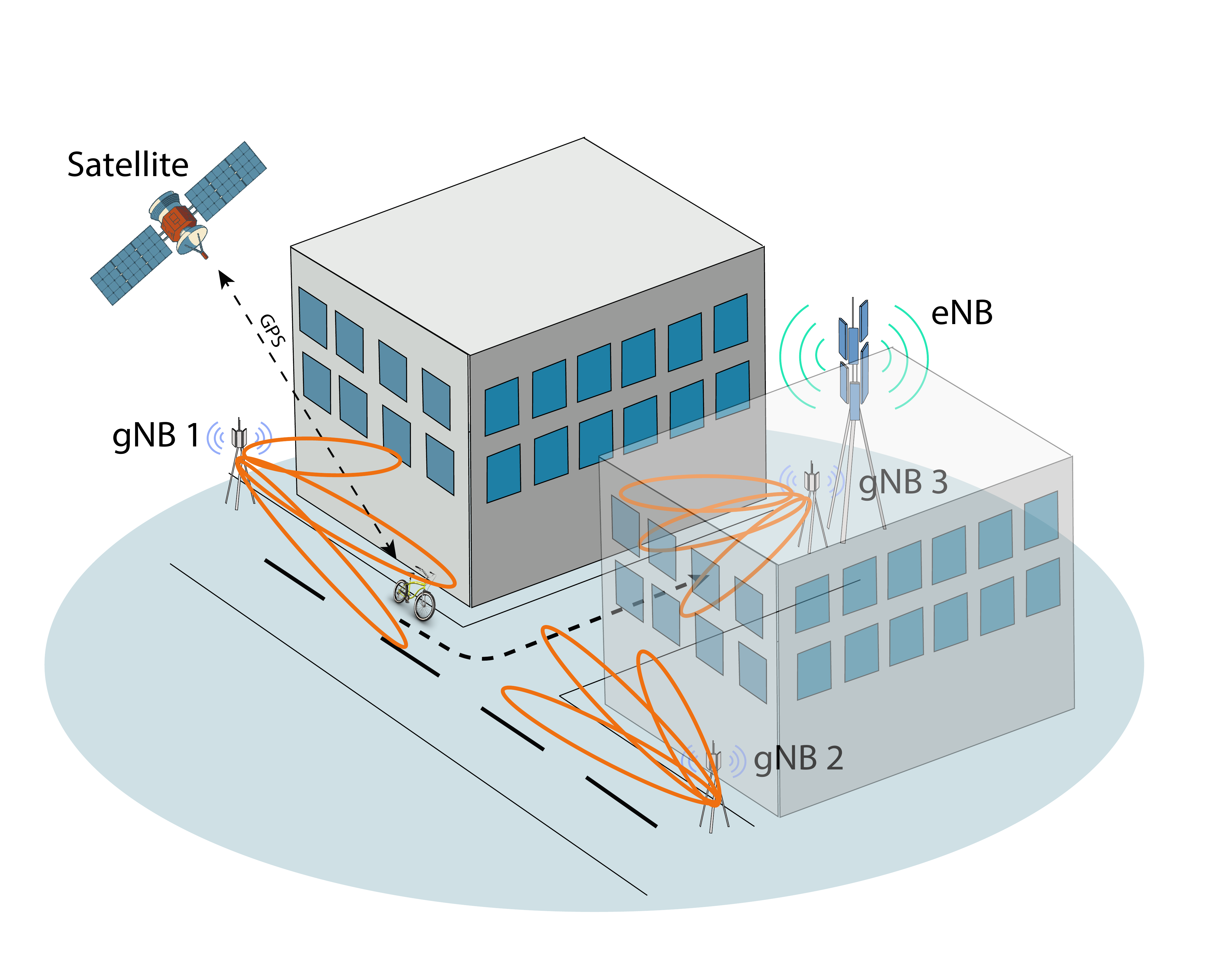}
  \setlength{\belowcaptionskip}{-14pt}
  \caption{Overview of the scenario: a UE uses GPS context information and dual connectivity to improve handover latency in an urban scenario.}
  \label{scenario}
\end{figure}

Typically, an RL agent optimizes its action based on a customized reward or objective function. The design of a reward function will intend to maximize or minimize certain metric of interest. In addition, a reward function could also be used in scenarios where our agent's goal is known. However, in goal-directed problems as in the majority of RL problems, sparse rewards are a significant challenge that affects the learning of robust value functions and thus, optimal action selection. Hierarchical reinforcement learning (hRL) \cite{Kulkarni2016} helps to solve the aforementioned problem by splitting the value function into two levels: a meta-controller and a controller. 

\begin{figure*}
\center
  \includegraphics[scale=0.6]{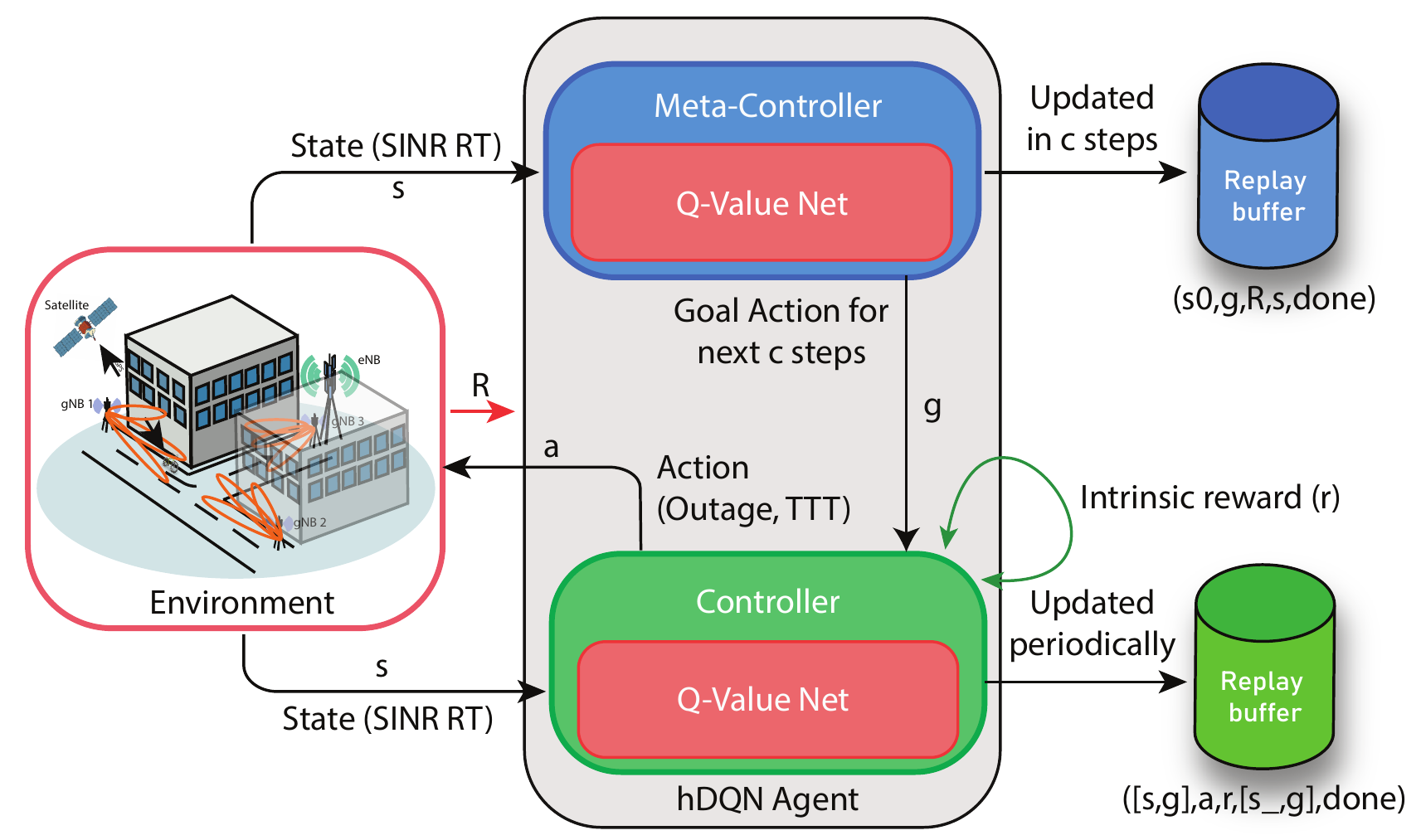}
  \setlength{\belowcaptionskip}{-12pt}
  \caption{Hierarchical Deep Q-Learning (HiDQL) applied in a Multiple Radio Access Technology (multi-RAT) LTE-NR scenario. The HiDQL agent located in the coordinator will optimize TTT and SINR outage metrics to reduce multi-RAT handover latency.} \label{scheme} 
  
\end{figure*}

In this paper, we address the handover problem in an LTE-NR network with dual connectivity using a hierarchical and a non-hierarchical architecture. Additionally, we consider the context information to further improve the DC algorithm performance.  
To do so, we use a novel hierarchical Deep Q-learning algorithm named hierarchical Deep Q-Learning (HiDQL) and a non-hierarchical RL approached named Clipped Double Q-Learning (CDQL) to  improve  handover latency in a DC architecture. HiDQL presents an architecture comprised by a meta-controller and a controller. Two parameters are adjusted in this problem: Threshold to Time (TTT) and  signal-to-interference-plus noise ratio (SINR) outage. TTT is defined as the time to hold the condition of triggering the handover algorithm named Secondary Cell Handover (SCH), discussed in section III. On the other hand, SINR outage is defined as the SINR threshold in which we consider a cell being in outage from the User Equipment (UE). The controller will propose actions in terms of  TTT and SINR outage aided by the proposed goals by the meta-controller in order to minimize handover latency. We compare our results with an algorithm presented in \cite{Rivera2021} named Clipped Double Q-Learning (CDQL) that proved its efficiency in load balancing problems. Our results show an improvement in terms of latency with a gain of 47.6\% and 26.1\% for Digital-Analog beamforming (BF),  17.1\% and 21.6\% for Hybrid-Analog BF, and 24.7\% and 39\% for Analog-Analog BF when comparing the RL-schemes with the baseline's best results. Additionally, we observed faster convergence when utilizing the CDQL algorithm, meanwhile the HiDQL showed improved results in terms of handover latency. Finally, we obtained an improvement regarding handover latency under line of sight assumptions when considering context information in comparison with no context available.

The rest of this paper is organized as follows. Section \ref{Section2} presents the existing works related to dual connectivity and hierarchical algorithms applications in wireless networks. System model is demonstrated in Section \ref{Section4}.
 Section \ref{Section5} provides a description of the Markov Decision process of the RL algorithms and baselines utilized. Section \ref{Section6} depicts the performance evaluation and comparison with the baseline solutions. Section \ref{Section7} introduces context information in the DC architecture and its performance evaluation. Finally, section \ref{Section8} concludes the paper.

\section{Related work} \label{Section2}
 \vspace{-1.5mm}
To the best of our knowledge, there is no previous work where hRL has been utilized to optimize handover key parameters such as TTT and SINR cell outage in a 5G dual connectivity scenario. The aforementioned parameters are defined in detail in section V. However, dual connectivity (DC) has been studied thoroughly in the literature. These studies are summarized below. 

In \cite{Yilmaz2019}, the authors give an overview of the 4G LTE-NR DC based on the new specifications given by 3GPP. A study case showed that DC is capable of providing coverage and capacity improvements when compared to standalone LTE-NR deployed individually. More recently, in \cite{Monserrat2020} the authors perform an analysis of the dual connectivity proposal in the 3GPP release 15\cite{3GPPTS37.3402019}. In addition to the existent specification, the authors remarked the possibilities to include a Secondary Cell controller to manage the multi-radio DC signaling. Furthermore, in \cite{Agiwal2021}, the authors present a survey on recent developments of 4G/5G DC, as well 4G/5G internetworking performance and future research challenges and open issues. 

Besides, the body of works in DC, the concept of hRL has not been explored much in the the wireless community.  In \cite{Geng2021}, the authors study outage avoidance in a two-hop cooperative relay network by utilizing hRL in optimizing relay selection and both source and relays transmission power. In \cite{Liu2021}, the authors propose a hierarchical deep Q-network to perform dynamic multi-channel spectrum sensing in cognitive networks. In this work the actions corresponds to the selection of channel and the reward corresponds to the channel status (busy/idle). 
In the following sections, we will introduce the dual connectivity architecture utilized in this work and explain how RL emerges as effective solution in the optimization of the handover in DC scenarios. Additionally, we forsee the advantage of embedding context-awareness in the DC architecture. 
\vspace{-1em}

\section{System model} \label{Section4}
\label{AA}
In this work, we use a DC architecture presented in \cite{Polese2017} inspired by a 3GPP's previous DC proposal \cite{3GPP}. In \cite{Polese2017}, the authors leverage the usage of a DC framework to improve, among others, handover latency by using an algorithm named Secondary Cell Handover (SCH). SCH enables fast switching between the LTE and 5G RATs and is managed by a defined entity named coordinator. In this architecture the eNB takes the role of coordinator and controls the multi-RAT switch. Two main parameters are controlled by the coordinator: TTT and SINR cell outage. As part of the SCH algorithm, the coordinator will trigger the SCH when any of the RATs involved report a better SINR and neither of the RATs are in outage. Additionally, it will check the condition to trigger the SCH algorithm for TTT seconds to avoid ping pong scenarios. This algorithm resembles the classical LTE and 5G handover with the exception that no initial access is necessary and no interaction with the MME is required thanks to the nature of the DC architecture. As mentioned, the handover will be triggered based on SINR and specifically using a table named the Complete Report Table (CRT) comprised by the UE's SINR report tables from each gNodeB (gNB). Each gNB will perform a sweep over a number of predefined directions and will sense the Sounding Reference Signals from the UE's to obtain SINR measurements. The collected data will be sent via X2 interface to the coordinator. The delay incurred to perform the measurements sweeps is directly related with the beamforming (BF) technology used by the UEs and gNBs. The aforementioned delay, $D$, is calculated as: 

\begin{equation}
    D = \frac{N_{gNB}N_{UE}T_{per}}{L},
    \label{delay_eq}
\end{equation}
where $N_{gNB}$ and $N_{UE}$ correspond to the number of required sweep directions  needed for measurement collection by the gNBs and UEs, respectively. $T_{per}$ is the SRS periodicity and $L$ corresponds to the BF capabilities that will present different values if the transceiver is fully digital or analog. The relationship between $D$ and $L$ can be calculated using typical values as depicted in Table \ref{Table1} (Modified from \cite{Polese2017}). 

\begin{table} [ht]
  \centering
  \caption{Relationship between $L$ and $D$}
  \label{Table1}
  \begin{threeparttable} 
  \begin{tabular}{c c c}  
    \toprule
    gNB-UE BF & $L$ (gNB/UE) & $D$\tnote{*} (ms)\\ [1ex] 
\hline 
Analog-Analog &  $1/1$ &   25.6  \\ 
Hybrid-Analog &  $2/1$ &  16.8 \\
Digital-Analog & $N_{gNB}/1$  & 1.6 \\ 
  \bottomrule
  \end{tabular}
  \begin{tablenotes}
  \item[*] D in Eq. \ref{delay_eq} is calculated with $T_{per}=200\mu{s}$, $N_{gNB}=16$ and $N_{UE}=8$
  \end{tablenotes}
  \end{threeparttable}
\end{table}

In this paper, we consider an LTE-NR network with dual connectivity and consisting of $M_T$ gNBs. Additionally, an eNB will act as a coordinator and thus, serving the $M_T$ gNBs. A mobile user is dual connected to one gNB and one eNB at the same time. We consider an urban scenario with two buildings as shown in Fig. \ref{scenario}. The channel considered for the eNB is the 3GPP Channel Model, meanwhile the 3GPP building channel type and losses based on the 3GPP UMi Street Canyon propagation model is used for the gNBs.

\section{Hierarchical Deep Q-Learning (HiDQL) } \label{Section5}

In this work, we use a state-of-the-art hierarchical Deep Q-learning (HiDQL) algorithm. This algorithm is presented in \cite{Kulkarni2016}, where the 
goal-directed behavior is studied under some specific sparse reward problems such as the Montezuma Revenge and a complex discrete stochastic decision process. 

As shown in figure \ref{scheme}, a hierarchical agent located in the coordinator observes the user's SINR report table and receives the extrinsic reward based on the feedback from the environment in terms of handover latency. This agent will adjust key parameters as TTT and SINR outage in order to maximize the agent's extrinsic reward. In the following subsection, we will present an overview of HiDQL and then formally define our solution.

\normalem 
\begin{algorithm}

\algsetup{linenosize=\tiny}
 \scriptsize

Initialize $c_{max}$, $\kappa$, experience replay buffers $\{B_{mc}$, $B_{c}\}$, policy networks $\{\mu_{mc}$, $\mu_{c}\}$, for the meta-controller and controller, respectively. 

\SetKwFunction{FMain}{Done}
\SetKwProg{Fn}{Function}{:}{}
\Fn{\FMain{$c$, $c_{max}$, $\mathcal{S}_\kappa$, $\kappa$ }}{

    \eIf{($c = c_{max}$ \textbf{or} $\mathcal{S}_\kappa < \kappa$)}
       {\textbf{return} True}
    {\textbf{return} False}}
 \textbf{End Function}

\For{environment step $t\gets1$ \textbf{to} $T$}{ 
    $c\gets0$\\
    Init environment and initial state\\
    Execute goal from meta-controller: $g=\mu_{mc}(s)$ \\
    \While {c$_{max}$ \textbf{not} reached} {
    $R\gets0$; $s_0\gets s$; $\mathcal{S}_\kappa$ = \textbf{false}\\
    
    \While {\textbf{not}  {\normalfont \texttt{Done}} } {
    Execute action from controller: $a=\mu_{c}(s,g)$\\
    Get next state $s'$ and similarity condition $\mathcal{S}_\kappa$ ;\\
    Calculate intrinsic reward $r$ and receive extrinsic reward $r_g$;\\
    Store $([s,g], a, r, [s',g], {\normalfont \texttt{Done}})$ in replay buffer $B_{c}$;\\
    Update $\mu_{c}$;\\
    $c:= c + 1$; $R:= R + r_g$; $s:=s'$\\
    
    }
    Store $(s_0, g, R, s,{\normalfont \texttt{Done}})$ in replay buffer $B_{c}$;\\
    Update $\mu_{mc}$;\\
    \If{\textbf{not} {\normalfont \texttt{Done}}}{
        Execute goal from meta-controller: $g=\mu_{mc}(s)$ 
     }   
    }
} 
\caption{Hierarchical Deep Q-Learning (HiDQL)}
 \medskip
 \label{hidql_algo}
\end{algorithm}
\vspace{-2em}
\subsection{Meta-controller and controller action space selection}\label{AA}
In the proposed hierarchical approach described in Algorithm \ref{hidql_algo}, the meta-controller proposes goals or high level actions when the maximum $c$ steps, $c_{max}$ is achieved or when the similarity between the goal and low-level action, $\mathcal{S}_\kappa$  falls under a predefined threshold index $\kappa$ as defined in the function \texttt{Done}. Meanwhile, the controller's low level actions are executed $c$ inner steps. For both levels, the actions are defined in the same fashion as:
\begin{equation}
    A(t) = \begin{bmatrix} a_{out}(t), a_{ttt}(t)\end{bmatrix} ,
    \label{action}
\end{equation}
where $a_{out}$ and $a_{ttt}$ are the SINR outage and TTT values, respectively. Such values are discretized  into $K_o$ and $K_{TTT}$ levels according the maximum and minimum defined values. The possible values for $a_{out}$ and $a_{ttt}$ are defined as:
\begin{equation}
    A_{out} = \{O_{min}, O_{min} + \frac{O_{max} - O_{min}}{K_{o}-1},..., O_{max}\},
\end{equation}
\begin{equation}
\begin{aligned}
    A_{ttt} = \{TTT_{min}, &TTT_{min} + \\ &\frac{TTT_{max} - TTT_{min}}{K_{TTT}-1}, 
    \cdots, TTT_{max}\},
\end{aligned}
\end{equation}
where $O_{min}$, $O_{max}$ and $TTT_{min}$, $TTT_{max}$ are predefined maximum and minimum values of SINR outage and TTT values, respectively. Finally, the size of the meta-controller and controller action spaces become $S = |A_{out}| * |A_{ttt}|$.

\begin{figure*}
\subfloat[ ]{\includegraphics[height=1.95in]{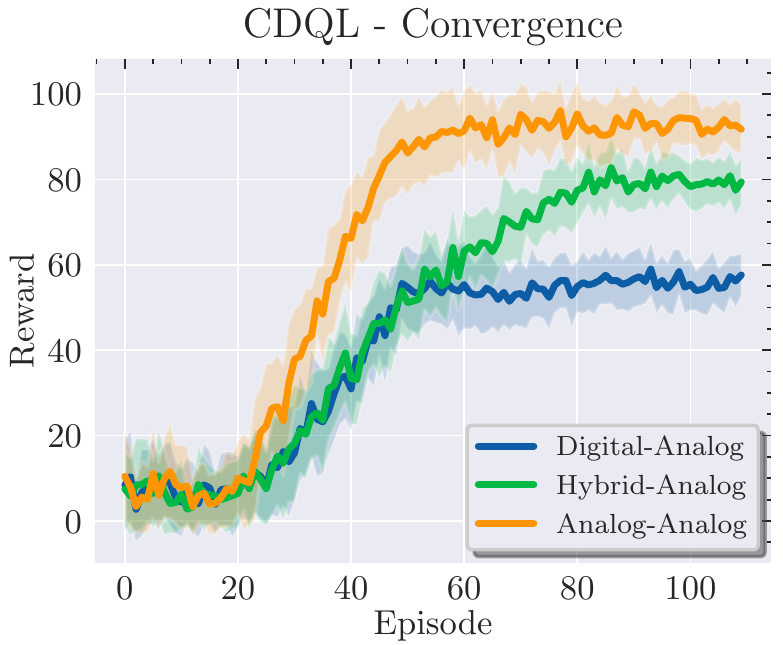}}
\subfloat[]{\includegraphics[height=1.95in]{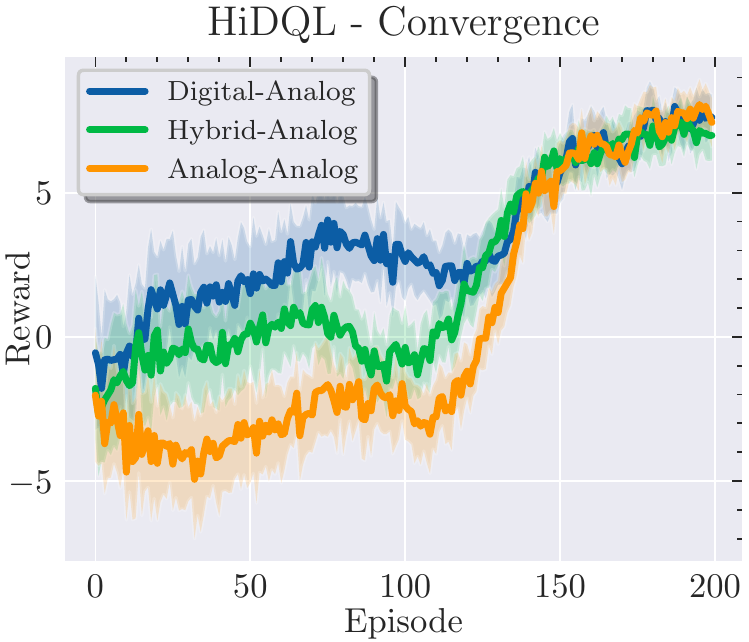}}
\subfloat[]{\includegraphics[height=1.95in]{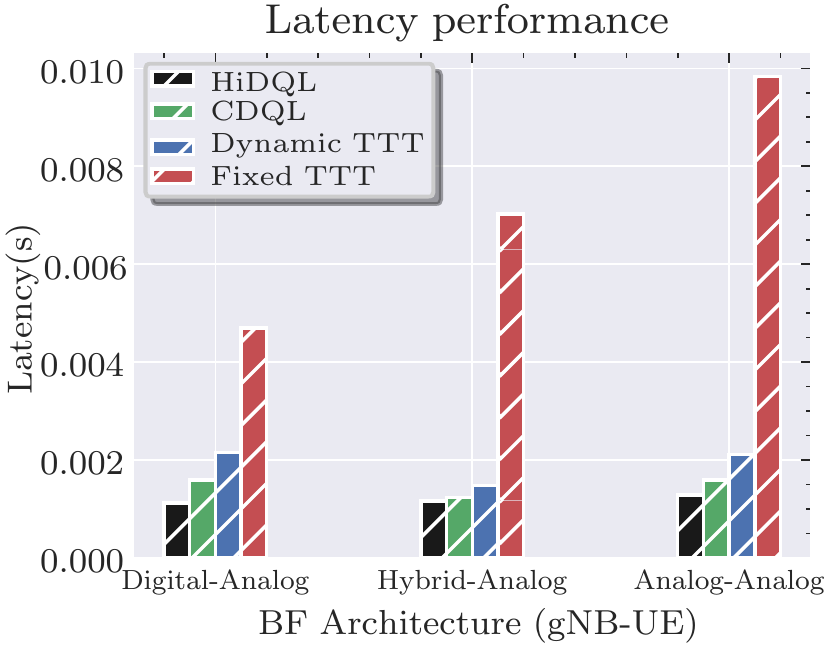}}
\setlength{\belowcaptionskip}{-12pt}
\caption{(a) CDQL convergence performance. (b) HiDQL convergence performance. contain. (c) Latency performance of the baselines Dynamic TTT and Fixed TTT and the proposed RL schemes CDQL and HiDQL.}
\label{results_a}
\end{figure*}

\begin{comment}

\begin{document}
    \begin{figure*}
\centering
\setkeys{Gin}{width=0.32\linewidth}
\includegraphics{example-image-duck}
\hfill
\includegraphics{example-image-duck}
\hfill
\includegraphics{example-image-duck}
    \caption{desired caption}
\label{fig_4}
    \end{figure*}
\end{document}